# FERMILAB SRF CRYOMODULE OPERATIONAL EXPERIENCE

A. Martinez, A. L. Klebaner, J. C. Theilacker, B. D. DeGraff, M. White and G. S. Johnson

Fermi National Accelerator Laboratory
Batavia, IL, 60510, USA

## ABSTRACT

Fermi National Accelerator Laboratory is constructing an Advanced Accelerator Research and Development facility at New Muon Lab. The cryogenic infrastructure in support of the initial phase of the facility consists of two Tevatron style standalone refrigerators, cryogenic distribution system as well as an ambient temperature pumping system to achieve 2K operations with supporting purification systems. During this phase of the project a single Type III plus 1.3 GHz cryomodule was installed, cooled and tested. Design constraints of the cryomodule required that the cryomodule individual circuits be cooled at predetermined rates. T hese constraints required special design solutions to achieve. This paper describes the initial cooldown and operational experience of a 1.3 GHz cryomodule using the New Muon Lab cryogenic system.

## INTRODUCTION

Fermi National Accelerator Laboratory is constructing an Advanced Accelerator Research and Development (AARD) facility at New Muon Lab. The AARD cryogenic system consists of two Tevatron style heat exchangers operating in a mixed mode – refrigeration and liquefaction. The refrigeration component is used for cooling of the low temperature thermal shields and to compensate for the distribution system heat leak, while the liquefaction component is used to cool the superconducting radio frequency (SRF) TESLA style cavities [1]. The high temperature thermal shields are cooled using liquid nitrogen. Ambient temperature vacuum pumping is used to reduce the helium vapor pressure in the cavities to achieve superfluid operation. A detailed system description was previously presented [2].

The cryogenic system was previously commissioned in various phases including individual components such as each refrigerator as well as the vacuum pump system. Subsequent commissioning tests were performed on the cryogenic distribution system using the single cavity, Capture Cavity 2 system (CC2), as the test load. More recently, the cryomodule Feedbox and Endcap manufactured by Cryotherm, GmbH were installed followed by the first ILC type cryomodule, designated CM1.

**CONTROLS**

The main control system consists of the Siemens Advanced Process Automation and Control System (APACS+$^{TM}$) which allows for re-configurable logic and loop control, as well as Input/Output (I/O) modules capable of handling a variety of signals. The control of localized equipment such as the compressors, expansion engine and vacuum pump are done using localized, self-contained commercial PLC based controls which communicate directly with the APACS system using a fiber optic line. Touch displays allow for local manipulation and control of these systems. A Fermilab designed Synoptic graphical user interface (GUI) is used to display and control all the parameters in a graphical format [3].

**CM1 COOLDOWN**

Cooldown of CM1 began on November 17, 2010. Total cooldown from 300 K to 4.5 K was achieved in 54 hours. Subsequent pumpdown to 1600 Pa was performed on November 22, 2010 and took approximately 3 hours. FIGURE 2 shows the progression of the cooldown of the various circuits from 300 K to 4.5 K.

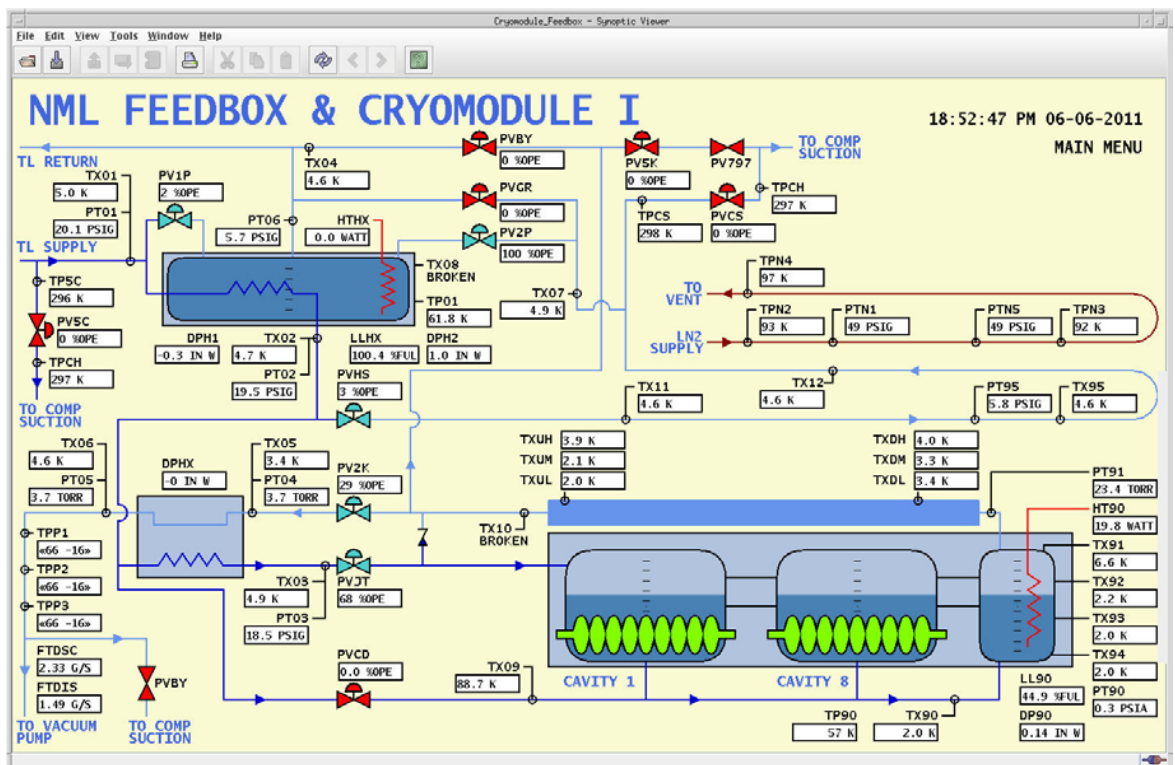

**FIGURE 1**. CM1 Graphical User Interface.

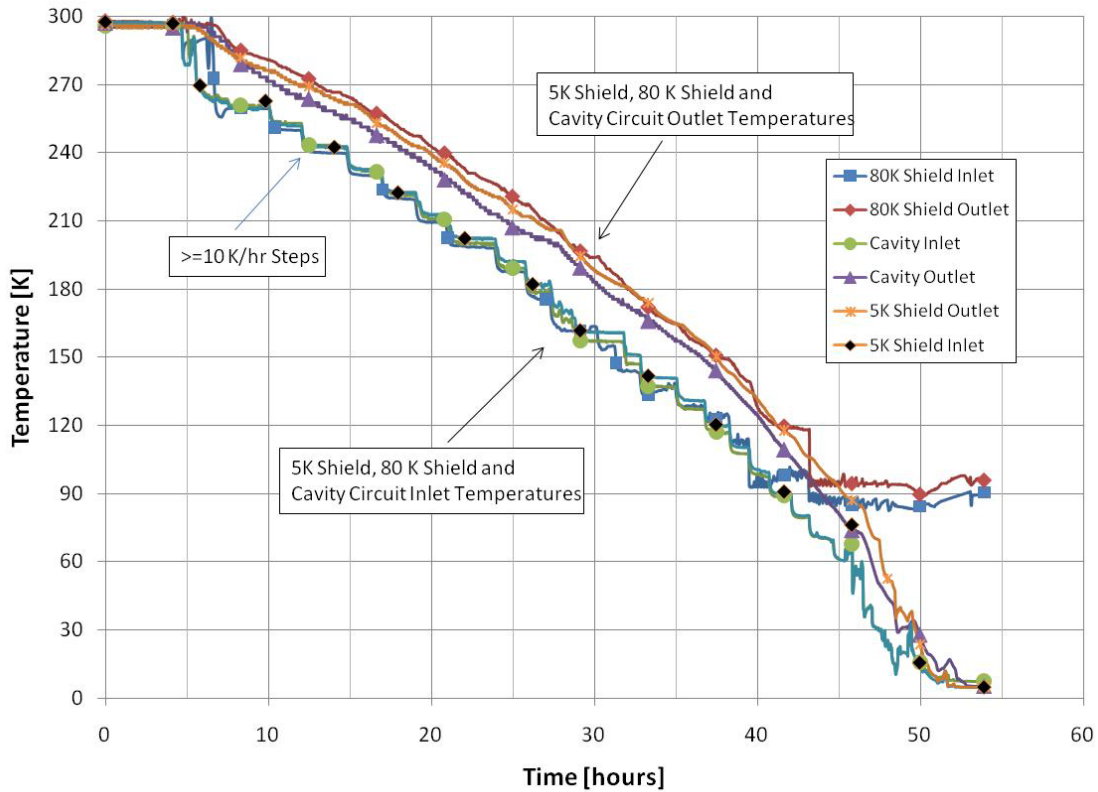

**FIGURE 2**. CM1 5K Shield, 80K Shield and Cavity Circuit Cooldown Temperatures.

The 1.3 GHz cryomodule design requires that its individual circuits be cooled at predetermined rates in order to limit thermal stresses. One such circuit is the helium gas return pipe (GRP) which is a 300 mm pipe that acts as the rigid support (strong back) from which all of the cavities are suspended. Based on a computer simulation of the cryomodule cooldown and operating experience with similar design modules at DESY, the maximum vertical gradient along this circuit is limited to less than 15 degrees. The longitudinal gradient is to be maintained at 50 degrees or less and lastly the overall cooldown rate is to be 10 degrees per hour or less over the range of temperatures from approximately 300 K to 100 K. The thermal shields circuits have identical longitudinal cooldown rate constraints.

FIGURE 3 shows that the maximum longitudinal temperature gradient requirements of the GRP and shield circuits were satisfied throughout the cooldown. For each circuit, the supply temperature was reduced by a maximum of 10 degrees each hour between the temperature range of 300 K to 100 K while monitoring the longitudinal and vertical temperature differences. In certain instances the step change was delayed until the temperature difference stabilized.

The gas return pipe was instrumented with a set of three in-flow Cernox® temperature sensors at each end of the GRP. The sensors were attached to a post in the center of the pipe and spaced to measure the temperature of the helium flow at the top, middle and bottom of the pipe. The difference between the top and bottom sensors determined the vertical gradient in the GRP. Additional surface mounted sensors were installed at each end to measure the surface temperature at the top, side and bottom of the pipe. These sensors were monitored during cooldown to maintain the desired vertical gradients. FIGURE 4 displays the vertical gradient, difference between the top and bottom temperatures, of the GRP during cooldown for both the in-flow and surface mounted sensors. Both the Endcap and Feedcap vertical gradients satisfied the 15 degree requirement throughout the cooldown. As expected, the data indicates that the in-flow sensors react more quickly with a better response.

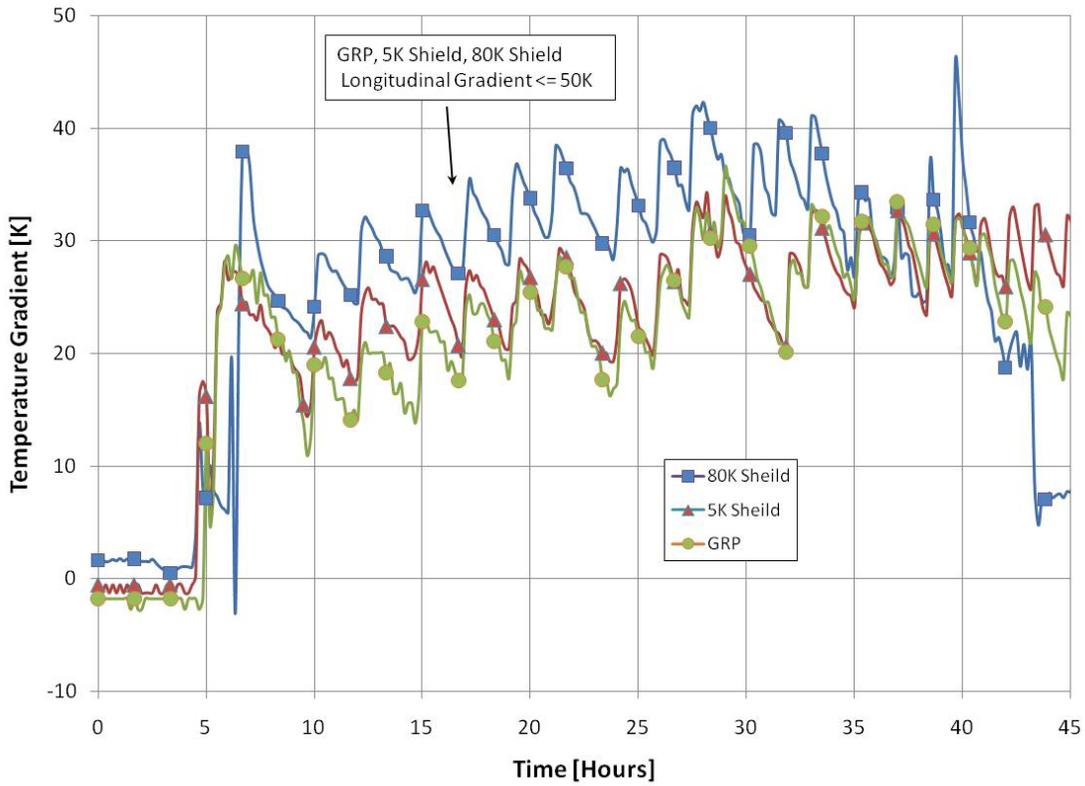

**FIGURE 3**. CM1 GRP, 5K Shield and 80K Shield Longitudinal Gradient Temperature.

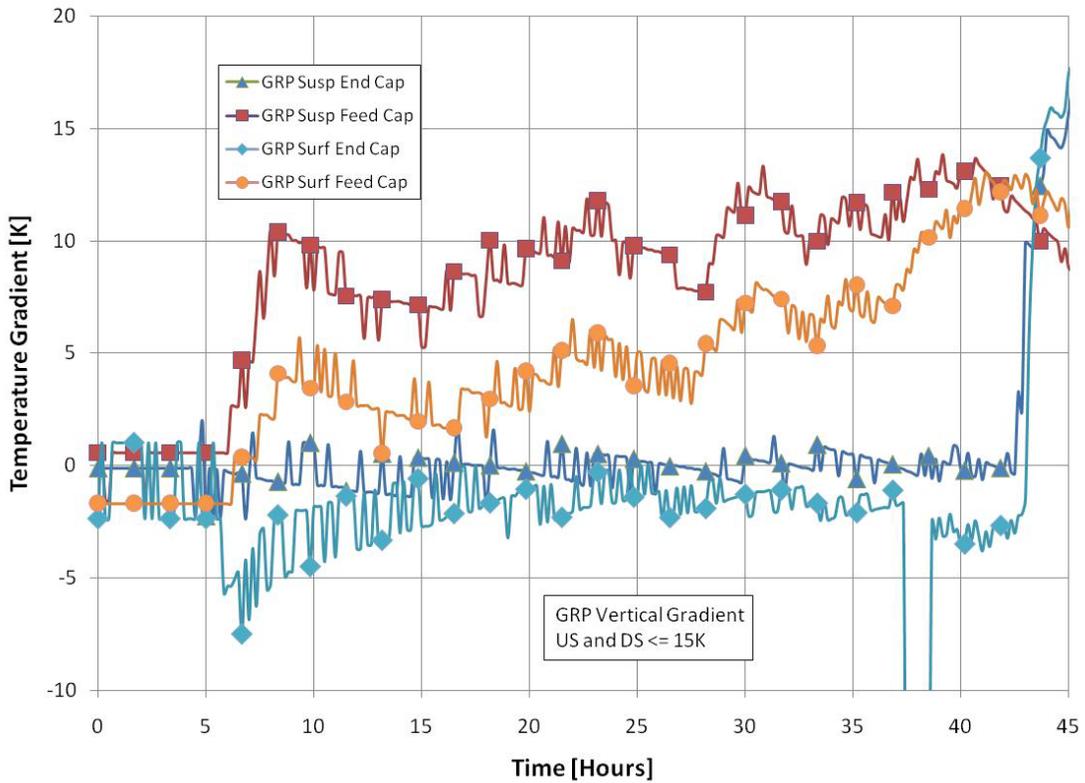

**FIGURE 4**. CM1 300 mm Gas Return Pipe Vertical Gradient Temperature.

The helium temperature entering the cryomodule cavity and low temperature shield circuit was controlled by coarsely controlling the temperature on the outlet of both refrigerators and fine tuning it using a specially designed mixing chamber upstream of the cryomodule Feedbox. In this way the cryomodule supply temperature could be accurately regulated at any temperature between 300 K and 100 K. During cooldown, the helium supply temperature was reduced in approximately 10 degree steps every hour. Once the 100 K temperature level was reached the cooldown rate constraints were released and the cooldown rate was increased to the maximum.

Since the high temperature thermal shield circuit was cooled using liquid nitrogen a different approach was necessary. To accurately control the temperature entering the cryomodule during cooldown, an inline 15 kW electrical vaporizer located upstream of the Feedbox was used to control the temperature of the supplied nitrogen. In order to maintain the proper cooldown rate, the outlet temperature of the vaporizer was also reduced in 10 degree steps every hour during cooldown. Once the 100 K temperature level was reached, the nitrogen vaporizer was turned off and the flow redirected.

## OPERATIONAL EXPERIENCE

Since the initial cooldown over six months ago, CM1 has remained at cryogenic temperatures and is undergoing radio frequency (RF) power testing. During RF power testing downtime, cryogenic system capacity tests were performed. The cryogenic system has a sustainable excess delivered capacity of approximately 110 W at superfluid temperatures. It is estimated that the static heat load of the cryomodule Feedcap, Endcap and CM1 is 18 W at superfluid helium temperatures. The accuracy of the estimate is 20%.

The system maintains cavity pressure well within specified requirements. Typically observed steady state cavity pressure fluctuation is +/- 10 Pa. Cavity liquid level is controlled using a dynamically adaptive PID control loop for precision level control [4]. FIGURE 5 shows the stability of the CM1 liquid level and pressure as an individual cavity is ramped to a gradient of over 25 MV/m. Only individual cavity testing is being done at this time.

## TECHNICAL ISSUES

While conducting the system cryogenic capacity tests, it was found that the mass flow meter used in the capacity measurement was under predicting the actual value. The flow was measured utilizing a thermal mass flow meter by FCI® (Model ST98) located on the discharge of the vacuum pump system. Calibration tests using a secondary variable area flow meter were performed. These tests indicated that the thermal mass flow meter was under-predicting the mass flow rate by as much as 30%. An electrical heater located in the Endcap phase separator was used to derive a mass flow across the full range of the flow meter.

FIGURE 6, shows the test data and comparison between the two flow measurements. Discussions with the flow meter manufacturer are currently underway to understand the discrepancy and to attempt to resolve the issue. Alternative flow meter types and designs are also being explored.

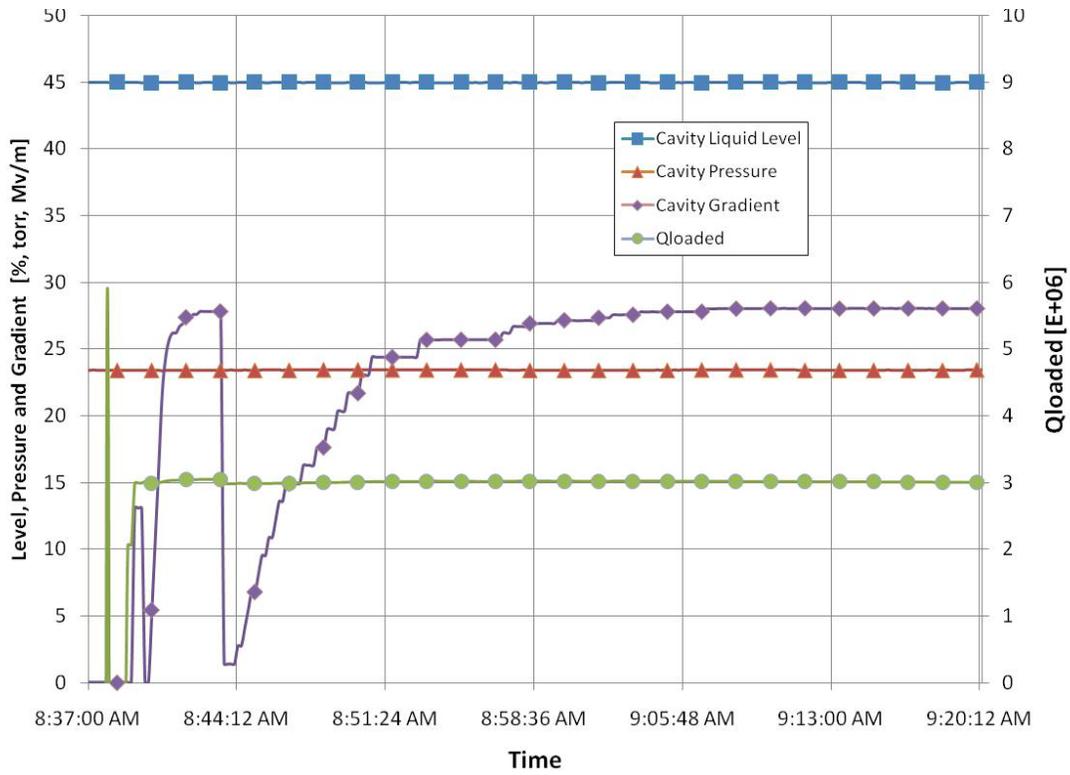

**FIGURE 5**. CM1 Cavity Level and Pressure Stability during Single Cavity RF Testing (Cavity #4).

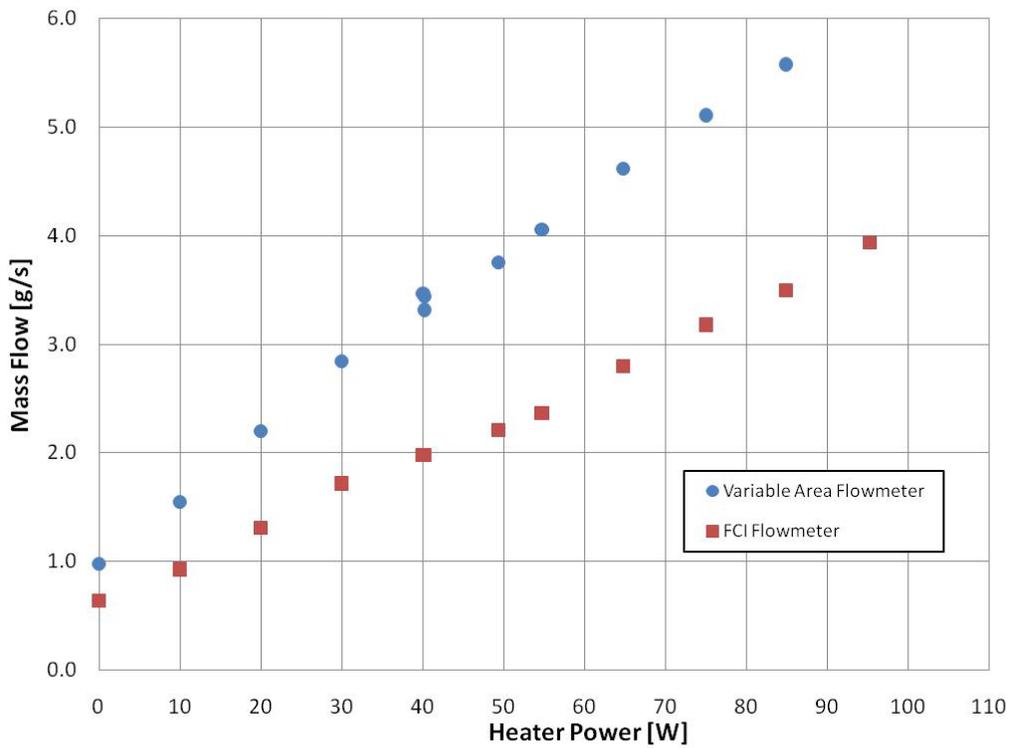

**FIGURE 6**. CM1 Thermal Flow Meter Testing.

## FUTURE PLANS

Future plans include improving compressor system availability by commissioning a purifier compressor and adding a spare main warm compressor. The purifier compressor takes the discharge gas from the vacuum system, compresses it and discharges it to the purifier/adsorber system before reentering the main system. The purpose behind this is that due to extended sub-atmospheric operation of the 2 K system there is always the possibility of air migration into the system. The purifier compressor would force this flow through the purifier/adsorber system. The addition of a spare warm compressor to the two existing compressors will provide added redundancy. The current configuration of operating with two refrigerators requires that two compressors operate continuously. The addition of a spare compressor will allow for maintenance activities on a single warm compressor while maintaining dual refrigerator operation.

## SUMMARY

The New Muon Lab cryogenic system has successfully supported initial CM1 cooldown and has been reliably operated for the past six months. The system has achieved the expected capacity and performance. Observations indicate that the static heat load of the CM1 system is on the order 18 W within 20% accuracy. Overall cryogenic system sustainable delivered capacity has been shown to be 110 W at superfluid temperatures. Cavity pressure regulation has been shown to be +/- 10 Pa with fine liquid level control using dynamically adaptive loop controls.

The New Muon Lab cryogenic system has unique capabilities specifically tailored for testing and operating of ILC cryomodules. The facility represents one of the two cryogenic systems in the world that allows for continuous operation of ILC cryomodules.

## ACKNOWLEDGEMENTS

Fermilab is operated by the Fermi Research Alliance, LLC under Contract No. DE-AC02-07CH11359. The authors wish to recognize the dedication and skills of the Accelerator Cryogenic Department technical personnel involved in the design, installation and testing of the cryogenic equipment.